\documentclass[prd,amsmath,showpacs,preprintnumbers,superscriptaddress,
twocolumn]{revtex4}
\usepackage{amsmath}
\usepackage{graphicx}

\newcommand{\beqa}{\begin{eqnarray}}
\newcommand{\eeqa}{\end{eqnarray}}

\newcommand{\oa}[1]{${\cal O}(a^{#1})$}

\begin{document}

\preprint{\vbox{
\rightline{ADP-03-142/T577}
\rightline{}}}

%\draft

\title{Unquenched Gluon Propagator in Landau Gauge}      

\author{Patrick O.~Bowman}
\affiliation{Special Research Centre for the Subatomic Structure of Matter 
and \\ The Department of Physics, University of Adelaide, SA 5005, 
Australia}
\affiliation{Nuclear Theory Center, Indiana University, Bloomington IN 47405, 
USA}
\author{Urs M.~Heller}
\affiliation{American Physical Society, One Research Road, Box 9000,
Ridge NY 11961-9000, USA}
\author{Derek B.~Leinweber}
\author{Maria B.~Parappilly}
\author{Anthony G.~Williams}
\affiliation{Special Research Centre for the Subatomic Structure of Matter 
and \\ The Department of Physics, University of Adelaide, SA 5005, 
Australia}

\date{February 17, 2004}

\begin{abstract} 
Using lattice quantum chromodynamics (QCD) we perform an unquenched
calculation of the gluon propagator in Landau gauge.  We use
configurations generated with the AsqTad quark action by the MILC
collaboration for the dynamical quarks and compare the gluon
propagator of quenched QCD ({\it i.e.}, the pure Yang-Mills gluon
propagator) with that of 2+1 flavor QCD.  The effects of the dynamical
quarks are clearly visible and lead to a significant reduction of the
nonperturbative infrared enhancement relative to the quenched case.
\end{abstract}

% PACS 12.38.Gc, 11.15.Ha, 12.38.Aw, 14.70.Dj 

%\vspace{1cm}
\maketitle

\section{Introduction}

The gluon propagator, the most basic quantity of QCD, has been subject
to much calculation and speculation since the origin of the theory.  In
particular there has long been interest in the infrared behavior of
the Landau gauge gluon propagator as a probe into the mechanism of
confinement~\cite{Mandula:1998tt}.  Some authors have argued it to be
infrared finite~\cite{Gribov:1978wm,Stingl:1986hx,Zwanziger:1991gz}
while others favored infrared singular~\cite{Mandelstam:1979xd,Buttner:1995hg}.
There is a long history of its study on the lattice, in quenched
QCD~\cite{Mandula:1987rh,Bernard:1994tz,Marenzoni:1995ap,Ma:1999kn,Becirevic:1999uc,Becirevic:1999hj,Nakajima:2000nr,Bonnet:2001uh,Langfeld:2001cz,Leinweber:1998uu,Leinweber:1998im,Bowman:2002fe,Bonnet:2000kw}
and in quenched $SU(2)$~\cite{Cucchieri:1998fy,Cucchieri:2003di}.
The restriction to quenched lattice gauge theory calculations has
been due to the lack of sufficient computational
resources.  The quenched theory differs from full QCD only in the
relative weighting of the background gauge configurations (due to the
fermion determinant), but the evaluation of the Green's functions is
otherwise the same.  In the quenched approximation the fermion
determinant is replaced by unity and this corresponds to the
complete suppression of all quark loops.  The removal of quark
loops is equivalent to the limit where all sea-quark masses are taken
to infinity.
In this paper, we report the first results for the gluon
propagator from an unquenched lattice computation.  

We study the gluon propagator in Landau gauge using configurations
generated by the MILC collaboration~\cite{Bernard:2001av} available
from the Gauge Connection ~\footnote{http://www.qcd-dmz.nersc.gov}.
These use ``AsqTad'' improved staggered quarks, giving us access to
relatively light sea quarks.  We find that the addition of dynamical
quarks preserves the qualitative features of the gluon dressing
function $q^2D(q^2)$ in the quenched case -- enhancement for
intermediate infrared momenta followed by suppression in the deep
infrared -- but produces a clearly visible effect.  A significant
suppression of the infrared enhancement with respect to the quenched
case is observed.  It is interesting to compare these results to those
of a recent Dyson-Schwinger equation study~\cite{Alkofer03}.

\section{Details of the calculation}

The gluon propagator is gauge dependent and we work in the Landau
gauge for ease of comparison with other studies.  It is also the
simplest covariant gauge to implement on the lattice.  Landau gauge is
a smooth gauge that preserves the Lorentz invariance of the theory, so
it is a popular choice.  It will be interesting to repeat this
calculation for the Gribov-copy free Laplacian gauge, but that will be
left for a future study.

The MILC configurations were generated with the \oa{2} one-loop
Symanzik improved \cite{Symanzik:1983dc} L\"{u}scher--Weisz gauge
action \cite{Luscher:1984xn}.  The dynamical configurations use the
``AsqTad'' quark action, an \oa{2} Symanzik improved staggered fermion
action.  $\beta$ and the bare sea-quark masses are matched such that
the lattice spacing is held constant.  The lattices we consider all
have the same dimensions.  This means that all systematics are fixed;
the only variable is the addition of quark loops.  The parameters are
summarized in Table~\ref{simultab}.  The lattice spacing is
approximately 0.125 fm~\cite{Davies:2003ik}.

\begin{table}[b!]
\caption{\label{simultab}Lattice parameters used in this study.  The
dynamical configurations each have two degenerate light quarks
(up/down) and a heavier quark (strange).  In physical units the bare
masses range from $\sim 16$ MeV to $\sim 79$ MeV.  The lattice spacing
is $a \simeq 0.125$ fm.}
\begin{ruledtabular}
\begin{tabular}{ccccccccc}
   & Dimensions      & $\beta$  &Bare Quark Mass  & \# Configurations \\            
\hline
1  & $20^3\times 64$ &   8.00   &quenched    & 192 \\            
2  & $20^3\times 64$ &   6.76   &0.01, 0.05  & 193 \\                         
3  & $20^3\times 64$ &   6.79   &0.02, 0.05  & 249 \\           
4  & $20^3\times 64$ &   6.81   &0.03, 0.05  & 212 \\       
5  & $20^3\times 64$ &   6.83   &0.04, 0.05  & 337 
\end{tabular}
\end{ruledtabular}
\end{table}

In Landau gauge the gluon propagator is entirely transverse.  In
Euclidean space, in the continuum, the gluon propagator has the tensor
structure
\begin{equation}
D_{\mu\nu}(q) = \left ( \delta_{\mu\nu}-\frac{q_{\mu}q_{\nu}}{q^2} \right) 
   D(q^2) \, ,
\label{eq:Landau-Prop}
\end{equation}
and at tree-level,
\begin{equation}
D(q^2) = \frac{1}{q^2}.
\end{equation}
With this lattice gauge action the propagator at tree-level is
\begin{equation}
D^{-1}(p_\mu) = \frac{4}{a^2}\sum_{\mu}\left\{ 
   \sin^2 \left( \frac{p_\mu a}{2} \right)
   + \frac{1}{3}\sin^4 \left( \frac{p_\mu a}{2} \right) \right\},
\label{eq:imp_tree}
\end{equation}
where 
\begin{equation}
p_\mu  = \frac{2 \pi n_\mu}{a L_\mu}, \qquad
n_\mu \in  \Bigl( -\frac{L_\mu}{2}, \frac{L_\mu}{2} \Bigr],
\label{eq:qhat}
\end{equation}
$a$ is the lattice spacing and $L_\mu$ is the length of the lattice in
the $\mu$ direction.  As explained in Ref.~\cite{Bonnet:2001uh}, this
suggests a ``kinematic'' choice of momentum,
\begin{equation}
q_\mu(p_\mu) \equiv \frac{2}{a}\sqrt{ 
   \sin^2 \left( \frac{p_\mu a}{2} \right)
   + \frac{1}{3}\sin^4 \left( \frac{p_\mu a}{2} \right) },
\end{equation}
ensuring that the lattice gluon propagator has the correct 
tree-level behavior.

The bare gluon propagator, $D(q)$ is related to the renormalized propagator
$D_R(q;\mu)$ through
\begin{equation}
D(q) = Z_3(\mu,a) \, D_R(q;\mu) \, 
\label{eq:renorm-def}
\end{equation}
where $\mu$ is the renormalization point.  In a renormalizable theory
such as QCD, renormalized quantities become independent of the
regularization parameter in the limit where it is removed.  $Z_3$ is
then defined by some renormalization prescription.  We choose the momentum
space subtraction (MOM) scheme where $Z_3(\mu,a)$ is determined by imposing
the renormalization condition
\begin{equation}
D_R(q)|_{q^2=\mu^2} = \frac{1}{\mu^2} \, ,
\label{eq:renorm-mom}
\end{equation}
{\it i.e.} it takes the tree-level value at the renormalization point.
In the following figures we have chosen $\mu = 4$ GeV.
%It follows that
%\begin{equation}
%q^2 D(q)|_{q^2=\mu^2} =  Z_3(\mu,a)
%\end{equation}
%at some chosen renormalization scale $\mu$.

\section{Simulations Results}

\begin{figure}[t]
\centering\includegraphics[height=0.99\hsize,angle=90]{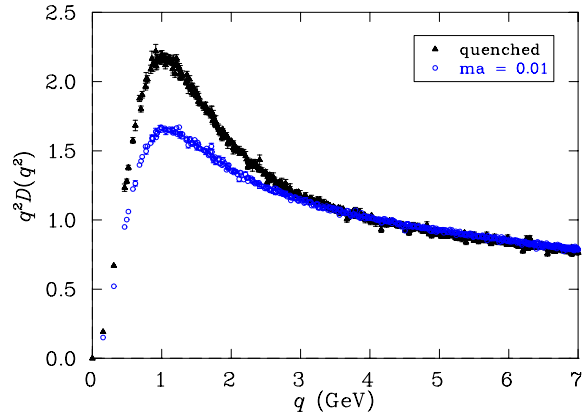}
\caption{Gluon dressing function in Landau gauge.  Full triangles
correspond to the quenched calculation, while open circles correspond
to 2+1 flavor QCD.  As the lattice spacing and volume are the same,
the difference between the two results is entirely due to the presence
of quark loops.  The renormalization point is at $\mu = 4$ GeV.  Data
has been cylinder cut \protect\cite{Leinweber:1998uu}.
\label{gp01}
}
\end{figure}

\begin{figure}[t]      
\begin{center}
   \includegraphics[height=0.99\hsize,angle=90]{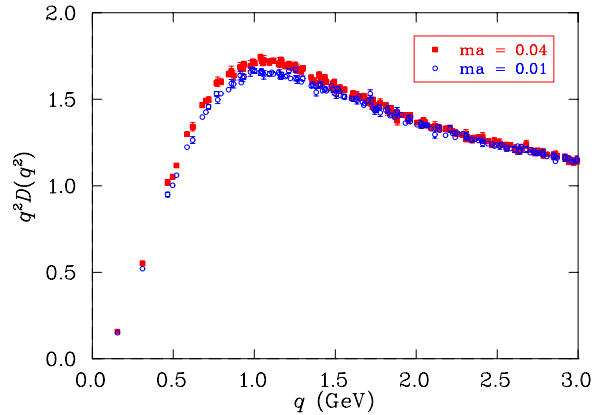}
\end{center}
\caption{The sea-quark mass dependence of the Landau gauge gluon
propagator dressing function renormalized at $\mu = 4$ GeV.  Filled
squares correspond to $u$ and $d$ bare masses $\simeq 63$ MeV and bare
$s$-quark mass $\simeq 79$ MeV.  Open circles correspond to the same
strange-quark mass, but with bare $u$ and $d$ masses $\simeq 16$ MeV.
Data has been cylinder cut \protect\cite{Leinweber:1998uu}.
Increasing the sea-quark masses alters the results in the expected
way, {\it i.e.}  towards the quenched data.
\label{gp02}
}
\end{figure}

Lattice studies strongly suggest that the quenched gluon propagator is
infrared finite~\cite{Bonnet:2001uh}.  As is customary, we will begin
by considering the (necessarily finite) gluon dressing function,
$q^2\, D(q^2)$.  In Fig.~\ref{gp01} we compare the well-known quenched
dressing function with that for 2+1 flavor QCD.  For the moment we
only consider the lightest of our dynamical quarks as we expect that
they will show the greatest difference from the quenched case.  

Indeed there is a clear difference between quenched and dynamical
quark behavior in the infrared region.  The addition of quark loops to
the gluon propagator softens the infrared enhancement without altering
its basic features. The screening of dynamical sea quarks brings the
$2+1$ flavor results significantly closer to the tree-level form,
$q^2\, D(q^2) = 1$.

In Fig.~\ref{gp02} we show the gluon dressing function for the
lightest and for the heaviest $u$ and $d$ quark masses in our set.
These correspond to bare light-quark masses of $\simeq 16$ MeV and
$\simeq 63$ MeV respectively; a factor of four difference.  The bare
strange-quark mass is the same in both cases ($\simeq 79$ MeV).  The
mass dependence of the gluon dressing function is only just
detectable.  We expect that increasing the sea-quark masses further
will interpolate between the curves in Fig.~\ref{gp01}.  We see that
the gluon propagator changes in the expected way.   As the sea-quark mass
increases, the curve moves toward the quenched result.  However,
for the range of bare quark masses studied here the change is
relatively small.  This transition would be better studied with
heavier sea quarks.

\begin{figure}[t]      
\centering\includegraphics[height=0.99\hsize,angle=90]{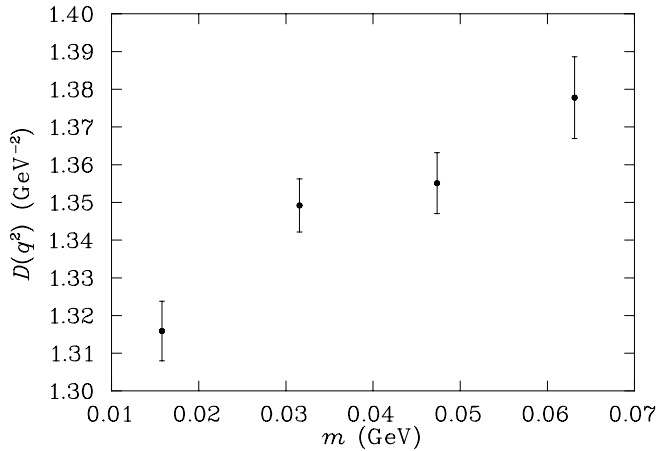}
\caption{The renormalized propagator at one momentum point in the
infrared hump of the gluon dressing function ($q \simeq 1.12$ GeV) is
shown here as a function of the bare light-quark mass.  }
\label{gp03}
\end{figure}

\begin{figure}[t]
\centering\includegraphics[height=0.99\hsize,angle=90]{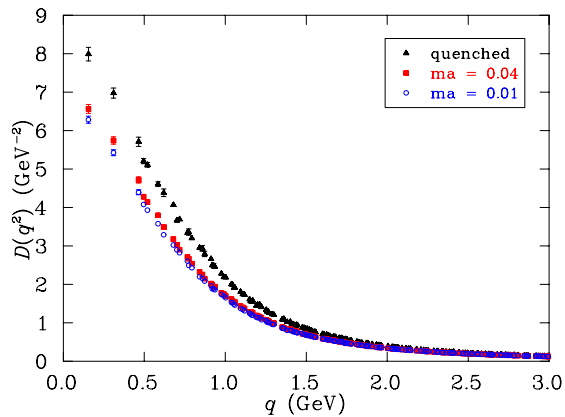}
\caption{The sea-quark mass dependence of the Landau gauge gluon
propagator renormalized at $\mu = 4$ GeV.  Filled triangles illustrate
the quenched propagator while filled squares correspond to bare
up/down masses $\simeq 63$ and bare strange-quark mass $\simeq 79$
MeV.  Open circles correspond to lighter bare up/down masses $\simeq
16$ MeV but with the same strange quark mass.  Data has been cylinder
cut \protect\cite{Leinweber:1998uu}.
\label{gp04}
}
\end{figure}

Another view of the mass dependence of the gluon propagator is
provided in Fig.~\ref{gp03}.  We choose one data point from the
infrared hump ($q \simeq 1.12$ GeV) and plot it for each choice of
bare light-quark mass.  Although the variation in the propagator at
this momentum is only 4.5\% over the range of quark masses investigated
here, the light sea-quark mass dependence is clearly resolved.

In Fig.~\ref{gp04} we present results for the gluon propagator,
$D(q^2)$.  The largest effects of unquenching are observed in the deep
infrared.  The shape of the curves suggest that previous results
indicating the infrared-finite nature of the quenched gluon
propagator~\cite{Bonnet:2001uh} are unchanged upon unquenching.  The
results suggest that the gluon propagator of QCD is infrared finite.
It will be interesting to examine the behavior of $D(0)$ as a function
of volume to elucidate this aspect of the gluon propagator further.

\begin{figure}[t]      
\centering\includegraphics[height=0.98\hsize,angle=90]{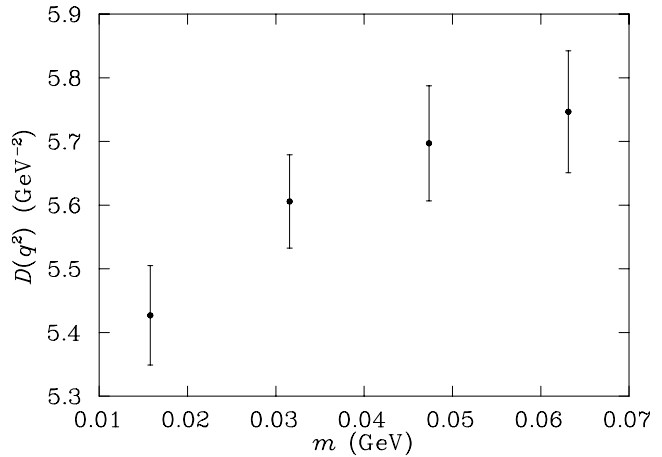}
\caption{The light sea-quark mass dependence of the renormalized gluon
propagator at a momentum point in the infrared region ($q \simeq 0.31$
GeV).  }
\label{gp05}
\end{figure}

Finally, in Fig.~\ref{gp05} the light sea-quark mass dependence of the
renormalized gluon propagator is illustrated for a momentum point in
the infrared region.  To avoid finite volume artifacts, the second
smallest nontrivial momentum is considered.  Whereas the mass
dependence of the propagator for the masses studied here is at the 4.5\%
level for $q \simeq 1.12$ GeV, the variance is larger in the infrared
region at 6\% for $q \simeq 0.31$ GeV.

\section{Conclusions}

The addition of quark loops has a clear, quantitative effect on the
gluon propagator.  While its basic structure is qualitatively similar
there is significant screening of the propagator in the infrared.  As
anticipated, the effect is to suppress the non-abelian enhancement of
the gluon propagator in the nonperturbative infrared-momentum region.
This is relevant to analytic studies of the gluon propagator and
confinement~\cite{Alkofer03}.  Despite the clear difference between
the quenched and dynamical results, we see little dependence on the
dynamical quark mass for the range of available light sea-quark
masses.  The dependence that is observed is consistent with
expectations.

Calculations on finer lattices are currently being made, which will
provide more information on the ultraviolet nature of the propagator
and provide a test for finite lattice spacing artifacts.  We would
like to extend the study to a wider range of dynamical masses to study
both the chiral limit and the transition to the quenched limit.
Finally, a study of the volume dependence of the propagator will
provide valuable insights into the nature of the propagator at
$q^2=0$.

\section*{ACKNOWLEDGMENTS}

This research was supported by the Australian Research Council and by
grants of time on the Hydra Supercomputer, supported by the South
Australian Partnership for Advanced Computing.

%%%%%%%%%%%%%%%%%%%%%%%%%%%%%%%%%%%%%%%%%%%%%%%%%%%%%%%%%%%%%%%%%%%%%%%%%%%%%%
%%%%%%%%%%%%%%%%%%%%% always use that to get your references    %%%%%%%%%%%%%%
\bibliographystyle{h-physrev}
%\nocite{*}
\bibliography{bib}
%%%%%%%%%%%%%%%%%%%%%%%%%%%%%%%%%%%%%%%%%%%%%%%%%%%%%%%%%%%%%%%%%%%%%%%%%%%%%%%
%%%%%%%%%%%%%%%%%%%%%%%%%%%%%%%%%%%%%%%%%%%%%%%%%%%%%%%%%%%%%%%%%%%%%%%%%%%%%%%

\end{document}